\documentclass[twocolumn,superscriptaddress,secnumarabic,
amssymb,amsmath,nobibnotes,aps,prd,showpacs,nofootinbib]{revtex4}%
\usepackage{graphicx}
\usepackage{epsf}
\usepackage{bm}
\usepackage{amsmath}
\usepackage{amsfonts}
\usepackage{amssymb}
\usepackage{epstopdf}
\usepackage{natbib}%
\usepackage{rotating}
\usepackage{pdflscape}
\usepackage{adjustbox}
\usepackage{hyperref}
\providecommand{\U}[1]{\protect\rule{.1in}{.1in}}
\newcommand{\be}{\begin{equation}}
\newcommand{\ee}{\end{equation}}

\newcommand{\mincir}{\raise
-3.truept\hbox{\rlap{\hbox{$\sim$}}\raise4.truept\hbox{$<$}\ }}
\newcommand{\magcir}{\raise
-3.truept\hbox{\rlap{\hbox{$\sim$}}\raise4.truept\hbox{$>$}\ }}

\usepackage{color}

\begin{document}

\title{Theoretical and observational bounds on some interacting vacuum energy scenarios}

\author{Weiqiang Yang}
\email{d11102004@163.com}
\affiliation{Department of Physics, Liaoning Normal University, Dalian, 116029, People's Republic of China}

\author{Supriya Pan}
\email{supriya.maths@presiuniv.ac.in}
\affiliation{Department of Mathematics, Presidency University, 86/1 College Street, Kolkata 700073, India}

\author{Llibert Arest\'e Sal\'o}
\email{l.arestesalo@qmul.ac.uk}
\affiliation{School of Mathematical Sciences, Queen Mary University of London, Mile End Road, London, E1 4NS, United Kingdom}

\author{Jaume de Haro}
\email{jaime.haro@upc.edu}
\affiliation{Departament de Matem\`atiques, Universitat Polit\`ecnica de Catalunya, Diagonal 647, 08028 Barcelona, 
Spain}

\begin{abstract}
The dynamics of interacting dark matter-dark energy models is characterized through an interaction rate function quantifying the energy flow between these dark sectors. In most of the interaction functions, the expansion rate Hubble function is considered and sometimes it is argued that, as the interaction function is a local property, the inclusion of the Hubble function may influence the overall dynamics. This is the starting point of the present article where we consider a very simple interacting cosmic scenario between vacuum energy and the cold dark matter characterized by various interaction functions originated from a general interaction function:  $Q=  \Gamma\rho _{c}^{\alpha }\rho _{x}^{1-\alpha -\beta}(\rho _{c}+\rho_{x})^{\beta}$, where $\rho_c$, $\rho_x$ are respectively the cold dark matter density and vacuum energy density; $\alpha$, $\beta$ are real numbers and $\Gamma$ is the coupling parameter with dimension equal to the dimension of the Hubble rate.  We investigate four distinct interacting cosmic scenarios and constrain them both theoretically and observationally. Our analyses clearly reveal that the interaction models should be carefully handled.   
\end{abstract}

\maketitle
\section{Introduction}
\label{sec:intro}

The theme of the present work is to consider a  generalized cosmic scenario where dark matter and dark energy, two heavy components of the universe, are interacting non-gravitationally.
Observational data suggest that nearly 96\% of the total energy density of the universe is occupied by these dark fluids. The dark matter sector is responsible for the structure formation of our universe, while, due to the presence of the dark energy fluid, the expansion of our universe is currently accelerated. The dynamics of both fluids is not clear and that is why various cosmological models have been proposed so far. The simplest cosmological scenario is the one where none of the fluids, especially dark matter and dark energy, interacts with each other apart from the gravitational interaction between them. The Concordance $\Lambda$-Cold-Dark-Matter ($\Lambda$CDM;  $\Lambda >0$ being the cosmological constant) and $w$CDM models are some of the models in this group, where $w$ is the Equation of State parameter of the dark energy fluid which for the cosmological constant assumes $w =-1$. On the other hand, one may allow a non-gravitational interaction between dark matter and dark energy, leading to a  more generalized cosmic scenario, known as interacting cosmological model.

The  question that naturally arises is the following: what drives us to consider the interacting cosmological theories? In other words, what are the limitations of the non-interacting cosmological models? To answer this question we recall two well-known problems associated with the non-interacting theories, namely the cosmological constant problem and the coincidence problem.  Thus, the investigations aiming to find the explanations of the above two problems were in progress by many investigators.  Before 1990, Wetterich proposed that the tiny value of the cosmological constant can be explained if we allow a coupled system between the gravity and a scalar field \cite{Wetterich:1994bg}. Around the end of the nineties, when a convincing picture of the accelerated expansion of our universe appeared and the need of dark energy was justified, Amendola found that Wetterich's interaction proposal can be generalized by allowing an interaction between the dark matter and dark energy in order to explain the coincidence problem \cite{Amendola:1999er}. Subsequently, Amendola's proposal was supported by other investigators \cite{Huey:2004qv,Cai:2004dk,Pavon:2005yx,delCampo:2008sr,delCampo:2008jx}. And, with such appealing investigations, the theory of interaction started getting attention in modern cosmology \cite{Barrow:2006hia,He:2008tn,Valiviita:2008iv, Gavela:2009cy,Koyama:2009gd,Majerotto:2009np,Chimento:2009hj,Gavela:2010tm,Honorez:2010rr,Pan:2013rha,Salvatelli:2013wra,Costa:2013sva,Yang:2014gza,yang:2014vza,Salvatelli:2014zta,Yang:2014hea,Pan:2012ki,Nunes:2016dlj,Yang:2016evp,Pan:2016ngu,Erdem:2016hqw,Sharov:2017iue,Cai:2017yww,Yang:2017yme,Yang:2017zjs,Mifsud:2017fsy,Yang:2017ccc,vandeBruck:2017idm,Pan:2017ent,Li:2018jiu,Yang:2018pej,vonMarttens:2018iav,Gonzalez:2018rop,Yang:2018qec,Martinelli:2019dau,Paliathanasis:2019hbi,Li:2019loh,Yang:2019bpr,Yang:2019vni,Barrow:2019jlm,Li:2019ajo,DiValentino:2019jae,vonMarttens:2019ixw,Lucca:2020zjb,DiValentino:2020kpf,vonMarttens:2020apn,Yang:2021hxg,Bonilla:2021dql,Johnson:2021wou,Kumar:2021eev}.

Gradually, it was found that interaction in the dark sector has many important outcomes. The presence of an interaction may push the dark energy equation of state to cross the phantom divide line \cite{Huey:2004qv,Wang:2005jx,Sadjadi:2006qb,Pan:2014afa}. This is an interesting outcome in this context, because in order to explain the phantom crossing we need to introduce the negative sign before the kinetic term, which eventually invites instabilities both at the classical and quantum levels. Additionally, 
the interaction theory gained further attention due to solving some cosmological tensions, specially the 
$H_0$ and $S_8$ tensions. The tensions in both $H_0$ and $S_8$ have been a very serious issue which signal for new physics in the dynamics of the universe. The theory of interaction plays a very positive role to alleviate both tensions, for the $H_0$ tension see for instance \cite{Kumar:2017dnp,DiValentino:2017iww,Kumar:2019wfs,Yang:2018euj,Yang:2018uae,Pan:2019jqh,Pan:2019gop,DiValentino:2019ffd,Pan:2020bur}, and for the $S_8$ tension see \cite{Pourtsidou:2016ico,An:2017crg,Kumar:2019wfs}. In this context, we refer to a recent review on the $H_0$ tension \cite{DiValentino:2021izs}, where the ability of various Interacting Dark Energy (IDE) models of solving the $H_0$ tension has been  presented.
Thus, from the existing literature, it is evident that the investigations with the  interaction models should be continued and the present work has thus been motivated along this direction.

In the present work we have considered a variety of interaction models which differ significantly from most of the existing interaction models in the literature. Usually, in most of the cases, the interaction function, $Q$, is chosen as an analytic function of the energy density of dark matter ($\rho_c$) and the energy density of the dark energy ($\rho_x$) along with an explicit presence of the Hubble factor ($H$) of the Friedmann-Lema\^{i}tre-Robertson-Walker universe: $Q = H \mathcal{F} (\rho_c, \rho_x)$. The inclusion of the Hubble factor is mainly motivated to solve the continuity equations of the dark sectors' energy densities; however, its appearance can be justified as explored in \cite{Pan:2020mst}.  On the contrary, some people argue that, as interaction is a local property, then the global expansion factor cannot be associated with it. On the other hand, if the expansion of the universe suddenly stops, then the interaction rate vanishes -- this also leads us to think whether the interaction models should explicitly depend on the Hubble function or not. The cosmic dynamics are very complicated and it is very difficult to select the actual form of the interaction function based on our current understanding. Therefore, in the present work we have thus taken an attempt to investigate the properties 
of the interaction models which do now allow the explicit presence of the Hubble factor. We have primarily investigated the behavior of various key cosmological parameters due to the presence of such interaction functions and then constrained them in light of the recently available cosmological datasets.

The article has been organized in the following way. In section \ref{sec-2} we have described the basic equations of the universe allowing an interaction between dark matter and dark energy components. Section \ref{sec-3}
is fully devoted to understand the qualitative features of the interaction model. This section actually clarifies which interaction model should be rejected from the list. After that, in section \ref{sec-data+analyses} we describe the observational data and the constraints on the accepted interaction scenarios. Finally, in section \ref{sec-discuss} we conclude the present work with a brief summary of the whole article.

\section{Revisiting the Interacting Universe}
\label{sec-2}

As usual we assume the spatially flat Friedmann-Lema\^{i}tre-Robertson-Walker line element given by $ds^2  = -dt^2 + a^2 (t)\left( dx^2 + dy^2 + dz^2\right)$
to begin with the analyses of the interacting dark energy models. Here, $a(t)$ (hereafter we shall symbolize it only as $a$ without writing the cosmic time $t$) is the expansion scale factor of the universe. We further consider that the gravitational sector of the universe follows the Einstein gravity and, in addition, the universe contains several cosmic fluids, such as radiation (photons+neutrinos), baryons, pressureless dark matter and a vacuum energy, where the last two components, namely pressureless dark matter and dark energy, are interacting with each other. Thus, one can mathematically express such cosmological scenario with a set of equations as follows:

\begin{eqnarray}\label{cons-tot}
\dot{\rho}_{\rm tot} + 3 H (p_{\rm tot} + \rho_{\rm tot}) = 0,
\end{eqnarray}
where $\rho_{\rm tot}$ and $p_{\rm tot}$ are respectively the total energy density and total pressure comprising all the concerned cosmic fluids. In particular, $\rho_{\rm tot} =  \rho_r + \rho_b + \rho_c+\rho_x$ and, similarly, $p_{\rm tot} =  p_r + p_b + p_c+p_x$. The Hubble function $H \equiv \dot{a}/a$, appearing in the above equation (\ref{cons-tot}), gives the constraint on the dynamics as

\begin{eqnarray}\label{Hubble-equation}
H^2 = \frac{1}{3M_{pl}^2}\; \rho_{\rm tot} = \frac{1}{3M_{pl}^2}\; (\rho_r +\rho_b +\rho_c +\rho_x),
\end{eqnarray}  
where $M_{pl}$ denotes the reduced Planck's mass.
Since pressureless and vacuum energy are the only interacting fluids and others do not take part in the interaction process, from (\ref{cons-tot}) one can obtain the following equations:

\begin{eqnarray}
&&\dot{\rho}_r + 4 H \rho_r  =0 \Leftrightarrow \rho_r = \rho_{r0}\; a^{-4},\label{cons-rad}\\
&&\dot{\rho}_b + 3 H \rho_b  = 0 \Leftrightarrow \rho_r = \rho_{b0}\; a^{-3},\label{cons-baryons}\\
&&\dot{\rho}_c + 3 H \rho_c = -Q (t),\label{cons-cdm}\\
&&\dot{\rho}_x = Q (t).\label{cons-vacuum}
\end{eqnarray}
Here $\rho_{r0}$, $\rho_{b0}$ are respectively the present values of $\rho_r$, $\rho_b$. 
Notice that we have used the familiar relations $p_r = \rho_r/3$, $p_b  = 0$, $p_c = 0$ and we introduce a new function $Q (t)$, which is known as the coupling function between pressureless dark matter and vacuum energy and it actually determines the matter flow between these fluids.  One can clearly realize that to find the dynamics of the universe, equivalently the scale factor $a$, it is enough to solve the above set of conservation equations and then plug all $\rho_i$'s into the Hubble equation (\ref{Hubble-equation}). This actually needs the functional form for $Q (t)$. In the present work we propose some of such $Q (t)$ models in order to investigate the dynamics of the universe via recent observational evidences.

Let us consider a very general interaction model
\begin{eqnarray} \label{general-model}
Q=  \Gamma\rho _{c}^{\alpha }\rho _{x}^{1-\alpha -\beta }(\rho _{c}+\rho_{x})^{\beta },
\end{eqnarray}  
where $\alpha$ and $\beta$ are real numbers. For specific values of $\alpha$ and $\beta$, we can have several interaction models as follows,

\begin{eqnarray}
&& Q =  \Gamma \rho_c,\label{model0}\\
&& Q = \Gamma \rho_x,\label{model1}\\
&& Q =  \Gamma (\rho_c +\rho_x),\label{model2}\\
&& Q = \Gamma \frac{\rho_c \rho_x}{\rho_c + \rho_x}, \label{model3}
\end{eqnarray}
where $\Gamma$ is the coupling parameter having the same dimension as the Hubble parameter. Thus, the quantity $\bar{\Gamma}=\Gamma/H_0$, where $H_0$ represents the present value of the Hubble parameter, is dimensionless and we shall use this quantity in this work
assuming that $|\bar\Gamma|\leq 1$.  For convenience, from now on, we denote the cosmological scenarios driven respectively by the interaction functions (\ref{model0}), (\ref{model1}), (\ref{model2}) and (\ref{model3}), as IVS0, IVS1, IVS2 and IVS3 (standing for Interacting Vacuum Scenarios).

The interaction in the dark sector also modifies the perturbation equations. Therefore, the analysis at the level of perturbations is essential to understand the interacting dynamics. To derive the equations one can follow either the synchronous gauge or the conformal Newtonian gauge. Here, we work with the synchronous gauge and the line element in this case follows \cite{Ma:1995ey}

\begin{eqnarray}
\label{perturbed-metric}
ds^2 = a^2(\tau) \left [-d\tau^2 + (\delta_{ij}+h_{ij}) dx^idx^j  \right], 
\end{eqnarray}
where $\tau$ denotes the conformal time and $\delta_{ij}$,  $h_{ij}$ are respectively the unperturbed and perturbed metric  tensors.  

Now, for the above metric, the perturbations equations for the DM component will take the form 

\begin{eqnarray}
\delta _{c}^{\prime } &=&-\left( \theta _{c}+\frac{h^{\prime }}{2}\right) +%
\frac{aQ}{\rho _{c}}\left( \delta _{c}-\frac{\delta Q}{Q}\right) , \\
\theta _{c}^{\prime } &=&-\mathcal{H}\theta _{c},  \label{eq:perturbation}
\end{eqnarray}%
where $\delta_c = \delta \rho_c/\rho_c$ is the density perturbations for the pressureless DM and $\theta_c$ is the velocity perturbations for the same quantity. Here, the 
prime attached to any quantity refers to its derivative with respect to the conformal time $\tau$; 
$\mathcal{H}= a^{\prime}/a$ denotes the conformal 
Hubble factor; $h = h^{j}_{j}$ denotes the trace of the metric perturbations $h_{ij}$. Here note that, as the dark energy is vacuum, the density perturbations for vacuum, $\delta_x = \delta \rho_x/\rho_x$, will vanish. 
The quantity $\delta Q/Q$ for the general model (\ref{general-model}) assumes the form 
\begin{eqnarray}
\frac{\delta Q}{Q}= \alpha \delta _{c}+\beta\frac{\rho _{c}\delta _{c}}{\rho _{c}+\rho _{x}}.
\end{eqnarray}%
Let us note that the global expansion factor $H$ does not appear in this interaction model, that is in eqn. (\ref{general-model}), so the perturbations equations are different compared to them. 
Finally, in the dark matter comoving frame, since the density perturbations for the dark energy vanish, from the residual gauge freedom in the synchronous gauge, the velocity perturbation will also vanish, that means, $\theta_c=0$ \cite{Wang:2014xca}.  So, adjusting this into the previous equation, we end up with
\begin{eqnarray}
\delta _{c}^{\prime } &=&-\frac{h^{\prime }}{2} +%
\frac{aQ}{\rho _{c}}\left( \delta _{c}-\alpha \delta _{c}-\beta\frac{\rho _{c}\delta _{c}}{\rho _{c}+\rho _{x}}\right).
\end{eqnarray}%

Now, using the specific value of the `power-parameters', namely $\alpha$ and $\beta$, in the generalized interaction function, one could easily calculate the perturbation equations for the specific interaction models shown in eqns. (\ref{model0}) -- (\ref{model3}).

\begin{itemize}

\item The interaction function $Q=\Gamma\rho_c$ of eqn. (\ref{model0}) corresponds to  $\alpha=1$ and $\beta=0$ of the reference model (\ref{general-model}). Thus, one could derive the quantity $\delta Q/Q=\delta_c$ 
and, consequently,  the perturbation equation for DM becomes 
\begin{eqnarray}
\delta _{c}^{\prime } &=&-\frac{h^{\prime }}{2}.
\end{eqnarray}%

\item The interaction function $Q=\Gamma\rho_x$ of (\ref{model1}) corresponds to $\alpha=\beta=0$ of the reference model (\ref{general-model}). One could derive that $\delta Q/Q = 0$ and, therefore, the perturbation equation for DM becomes 
\begin{eqnarray}
\delta _{c}^{\prime } &=&-\frac{h^{\prime }}{2}+\frac{aQ}{\rho _{c}}\delta_c=-\frac{h^{\prime }}{2}+\frac{a\Gamma\rho_x}{\rho_c}\delta_c.
\end{eqnarray}%

\item  The interaction function $Q=\Gamma (\rho_c + \rho_x)$ of (\ref{model2}) corresponds to   $\alpha=0$ and $\beta=1$ of the reference model (\ref{general-model}). Thus, the quantity $\delta Q$ for this model becomes
\begin{eqnarray}
\frac{\delta Q}{Q}=\frac{\rho _{c}\delta _{c}}{\rho _{c}+\rho _{x}}
\end{eqnarray}%
and, consequently, the perturbation equation for DM is 
\begin{eqnarray}
&&\delta _{c}^{\prime } =-\frac{h^{\prime }}{2}+\frac{aQ}{\rho _{c}}\left(\delta_c-\frac{\rho _{c}\delta _{c}}{\rho _{c}+\rho _{x}}\right) \nonumber\\ 
&& =-\frac{h^{\prime }}{2}+\frac{a\Gamma\rho_x}{\rho_c}\delta_c.
\end{eqnarray}%

\item Finally,  the interaction function 
$Q=\Gamma\rho_c\rho_x/(\rho_c+\rho_x)$ of (\ref{model3}) corresponds to the values $\alpha=\beta=1$ of the generalized interaction function (\ref{general-model}). In this case
\begin{eqnarray}
\frac{\delta Q}{Q}=\delta _{c}+\frac{\rho _{c}\delta _{c}}{\rho _{c}+\rho _{x}}
\end{eqnarray}%
and, hence, the perturbation equation for DM is
\begin{eqnarray}
&&\delta _{c}^{\prime } =-\frac{h^{\prime }}{2}+\frac{aQ}{\rho _{c}}\left(\delta_c-\delta_c-\frac{\rho _{c}\delta _{c}}{\rho _{c}+\rho _{x}}\right)\nonumber\\ 
&&=-\frac{h^{\prime }}{2}-\frac{a\Gamma\rho_c\rho_x}{(\rho_c+\rho_x)^2}\delta_c.
\end{eqnarray}%

\end{itemize}

Now, in the next section we shall examine the cosmological parameters influenced by the above interaction models. This will give a clear idea on the corresponding interaction scenarios.

\section{Qualitative analyses of the interaction models}
\label{sec-3}

Here, we mainly discuss the qualitative nature of the interaction models by investigating  their impacts on various cosmological parameters. In order to understand 
the evolution of the interacting scenarios at the background level, one can safely neglect $\rho_b$ and $\rho_r$ at late times, since the contributions from these sectors will not change the future dynamics of the universe. With such minimal approximation and using the definition of the density parameter for the $i$-th fluid,  $\Omega_i=\frac{\rho_i}{3H^2 M_{pl}^2}$, one can obtain the following dynamical system,
 \begin{eqnarray}\label{dyn}
  \left\{\begin{array}{ccc}
    \dot{\Omega}_c   &=& -3H\Omega_c\Omega_x-\frac{Q}{3H^2M_{pl}^2}, \\
     \dot{\Omega}_x  & =&   3H\Omega_c\Omega_x+\frac{Q}{3H^2M_{pl}^2},
  \end{array}\right.
  \end{eqnarray}
  with $\Omega_c+\Omega_x=1$. This  dynamical system is very crucial because it helps us to understand the dynamics of the models at late times. One may note that for $\Gamma  =0$, for any interaction model prescribed in this work, one could trace back the non-interacting cosmological model. In fact, taking as a time $N=\ln a$, the dynamical system representing the non-interacting case becomes 
 \begin{eqnarray}
  \left\{\begin{array}{ccc}
    {\Omega}'_c   &=& -3\Omega_c\Omega_x, \\
     {\Omega}'_x  & =&   3\Omega_c\Omega_x,
  \end{array}\right.
  \end{eqnarray}
 which is an autonomous dynamical system with the following two fixed points: (i) $(\Omega_c=0,\Omega_x=1)$, which is an attractor, and (ii)
 $(\Omega_c=1,\Omega_x=0)$, which is a repeller. Therefore, we obtain a late time  accelerating universe with $w_{\rm eff}=-1$.

An autonomous dynamical system equivalent to  (\ref{dyn}) is 
 \begin{eqnarray}\label{dynamical}
\left\{ \begin{array}{ccc}
\dot{\Omega}_c &=& -3H\Omega_c(1-\Omega_c)-\frac{Q}{3H^2M_{pl}^2}, \\
\dot{H} &=& - \frac{3}{2}H^2\Omega_c,
\end{array}\right.
\end{eqnarray}
and, using the dimensionless variables 
$\bar{t}=H_0t$, $\bar{H}=H/H_0$ and $\Gamma=\Gamma/H_0$, the system (\ref{dynamical}) becomes
\begin{eqnarray}\label{dynamical1}\left\{\begin{array}{ccc}
    {\Omega}'_c   &=& -3\bar{H}\Omega_c(1-\Omega_c)-\bar{Q}, \\
    \bar{H}'  & =&-\frac{3}{2}\bar{H}^2\Omega_c,
  \end{array}\right.
  \end{eqnarray}
where $\bar{Q}$ will be different for different interaction functions given in (\ref{model0}) -- (\ref{model3}). In fact, for the interaction function (\ref{model0}), (\ref{model1}), (\ref{model2}) and (\ref{model3}), the modified interaction function $\bar{Q}$ will respectively take the forms: 
\begin{eqnarray}
&& \bar{Q} =  \bar\Gamma \Omega_c,\label{modelbar0}\\
&& \bar{Q} = \bar\Gamma \Omega_x,\label{modelbar1}\\
&& \bar{Q} =  \bar\Gamma,\label{modelbar2}\\
&& \bar{Q} = \bar\Gamma {\Omega_c \Omega_x}. \label{modelbar3}
\end{eqnarray}
Thus, one could understand the dynamical behavior of the models for different functional forms of $\bar{Q}$.

 \subsection{IVS0}
 
\begin{figure*}
    \centering
    \includegraphics[width=1.0\textwidth,trim={0 3.5cm 0 3.5cm},clip]{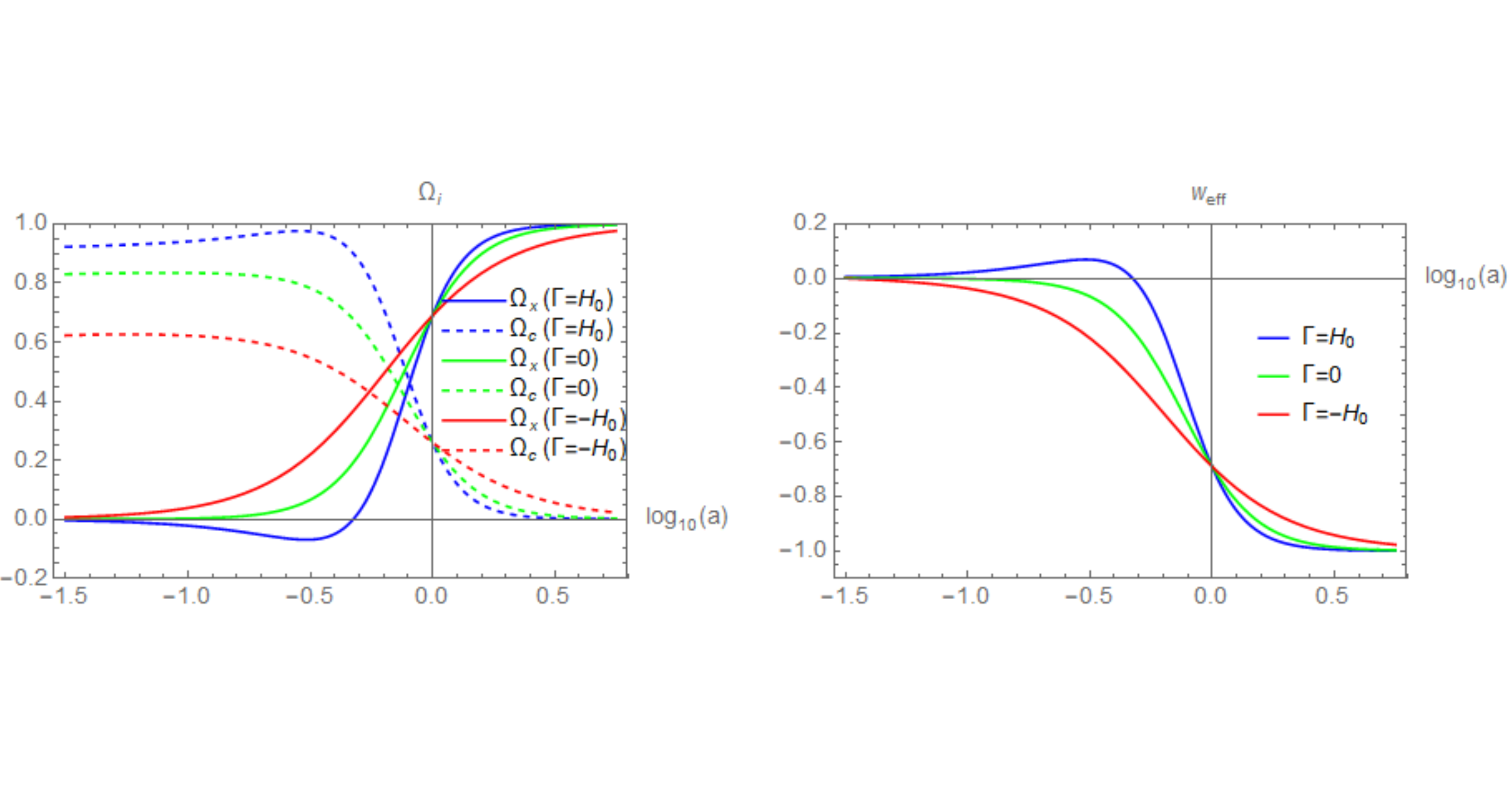}
    \caption{We show the evolution of the density parameters for dark matter and dark energy (left plot) and the total equation of state for different values of the dimensionless coupling parameter, namely $\Gamma/H_0= 1 $, $\Gamma/H_0=0$ and $\Gamma/H_0=-1$ for the interaction function (\ref{model0}). While drawing the plots, we have fixed $H_0=68$ km/sec/Mpc, $\Omega_{c0}=0.26$, $\Omega_{b0} = 0.0499$, $\Omega_{r0} = 0.0001$ and  $\Omega_{x0}=0.69$. }
    \label{fig:IVS0}
\end{figure*}

 At the background level there are two completely different situations separated by the non-interacting case $\Gamma/H_0=0$. We see in Fig. \ref{fig:IVS0} that for $\Gamma/H_0>0$ the model is not viable because the dark energy density $\rho_x$ becomes negative before the present time.

 In addition, in Fig. \ref{fig:IVS0} we show that the effective Equation of State (EoS) parameter $w_{\rm eff}$ converges to $-1$ at late times, obtaining an eternal cosmic acceleration.
 
 Analytically, we could show this behavior  using the eq. (\ref{dynamical1}), which for our model becomes
 \begin{eqnarray}
\left\{ \begin{array}{ccc}
{\Omega}'_c &=& -(3\bar{H}(1-\Omega_c)+\bar{\Gamma}) \Omega_c,\\
\bar{H}' &=&  -\frac{3}{2}\bar{H}^2\Omega_c.
\end{array}\right.
\end{eqnarray}
From the above two equations, one can write that 
\begin{eqnarray}
\frac{d\Omega_c}{d\bar{H}}=\frac{2\Omega_x}{\bar{H}}+
\frac{2\bar\Gamma}{3\bar{H}^2},
\end{eqnarray}
which allows us to obtain the analytical solution $\Omega_c(\bar{H})=\frac{2\bar\Gamma}{3\bar{H}}+\frac{C}
{\bar{H}^2}+1$, where  $C=-\frac{2\bar\Gamma }{3}-\Omega_{x,0}$.

As we will immediately see, the important case is  $C<0$, which implies $\bar\Gamma>-1.05$. But this always holds because we have already assumed $-1 < \bar\Gamma$.

Now we have to study the autonomous one-dimensional differential equation
$\bar{H}' =  -\frac{3}{2}\bar{H}^2\Omega_c$, which has as fixed points  $H=0$ and
\begin{eqnarray}\Omega_c=0\Longrightarrow \bar{H}_{\pm}=-\frac{\bar\Gamma}{3}\pm
\sqrt{\frac{\bar\Gamma^2}{9}-C},
\end{eqnarray}
because as $C$ is negative $\bar{H}_{\pm}$ are real numbers.
Note also that $\bar{H}_-$ is negative and, so, it could be disregarded. On the other hand,  $\bar{H}_+$ is an attractor.

Finally,
note that $\bar{H}'$ is positive when $\bar{H}\in (\bar{H}_-,\bar{H}_+)$, which implies that $\Omega_c$ 
is negative in this region, leading to a  non viable model.
Therefore,
we must impose $1>\bar{H}_+\geq 0$ ($\bar{H}=1$ is the initial condition), which always holds for  
 $\bar\Gamma>-3$, i.e., always happens. In conclusion, the system goes to $\Omega_c=0$, obtaining an accelerating  universe with effective EoS parameter $w_{\rm eff}=-1$.

 \subsection{IVS1}
\begin{figure*}
    \centering
    \includegraphics[width=1.0\textwidth,trim={0 3cm 0 3cm},clip]{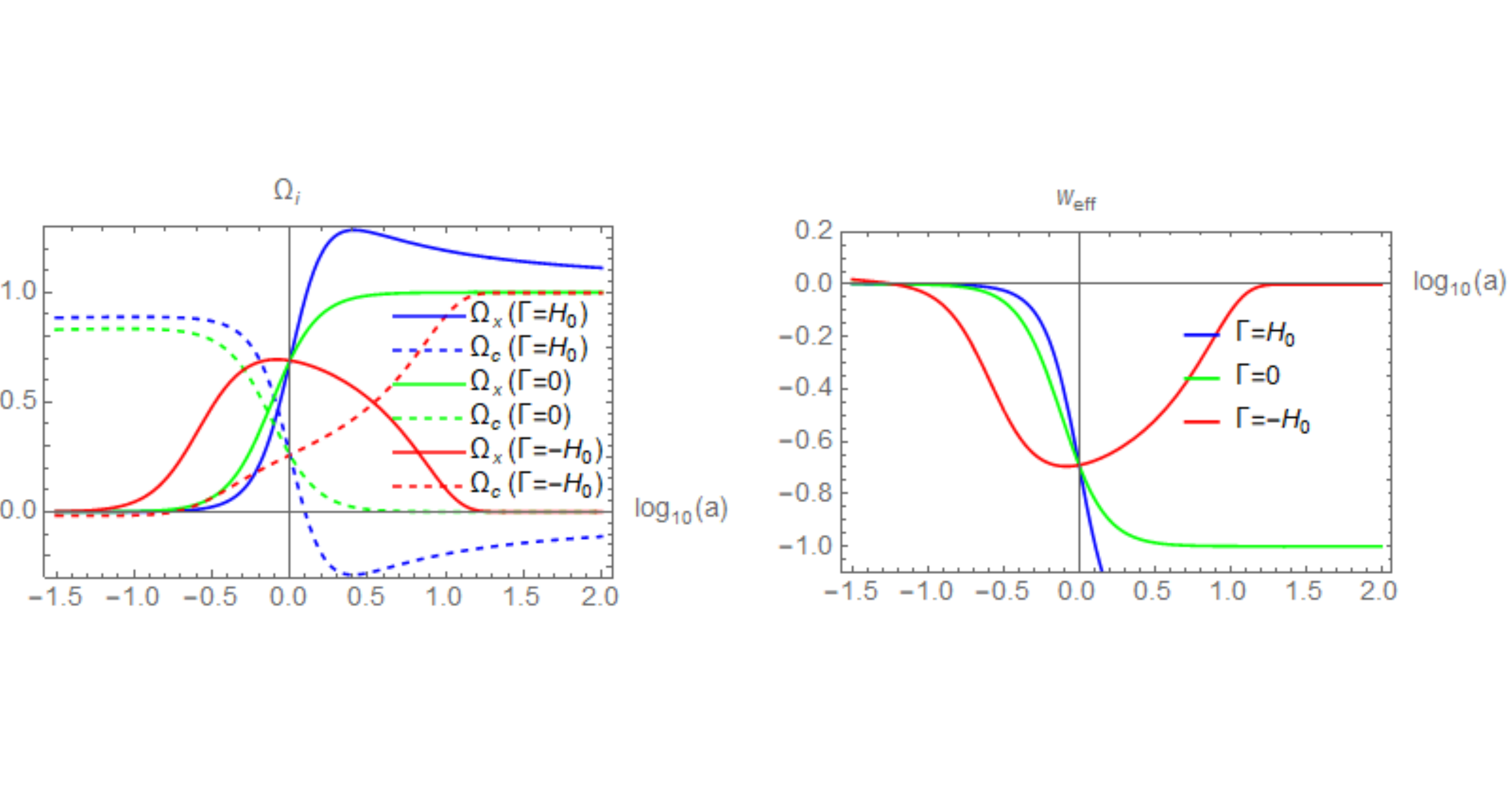}
    \caption{We show the evolution of the density parameters for dark matter and dark energy (left plot) and the total equation of state for different values of the dimensionless coupling parameter, namely  $\Gamma/H_0= 1 $, $\Gamma/H_0=0$ and $\Gamma/H_0=-1$ for the interaction function (\ref{model1}). While drawing the plots, we have fixed $H_0=68$ km/sec/Mpc, $\Omega_{c0}=0.26$, $\Omega_{b0} = 0.0499$, $\Omega_{r0} = 0.0001$ and  $\Omega_{x0}=0.69$.}
    \label{fig:IVS1}
\end{figure*}

  Once again at background level, for this model the case $\Gamma/H_0>0$ is also non viable because now the energy density of the cold dark matter $\rho_c$ becomes negative after the present time (see Fig. \ref{fig:IVS1}). But, for the viable negative case $\Gamma/H_0<0$, we see that at very late times $\rho_c$ dominates once again.

\begin{figure*}
\includegraphics[width=0.45\textwidth]{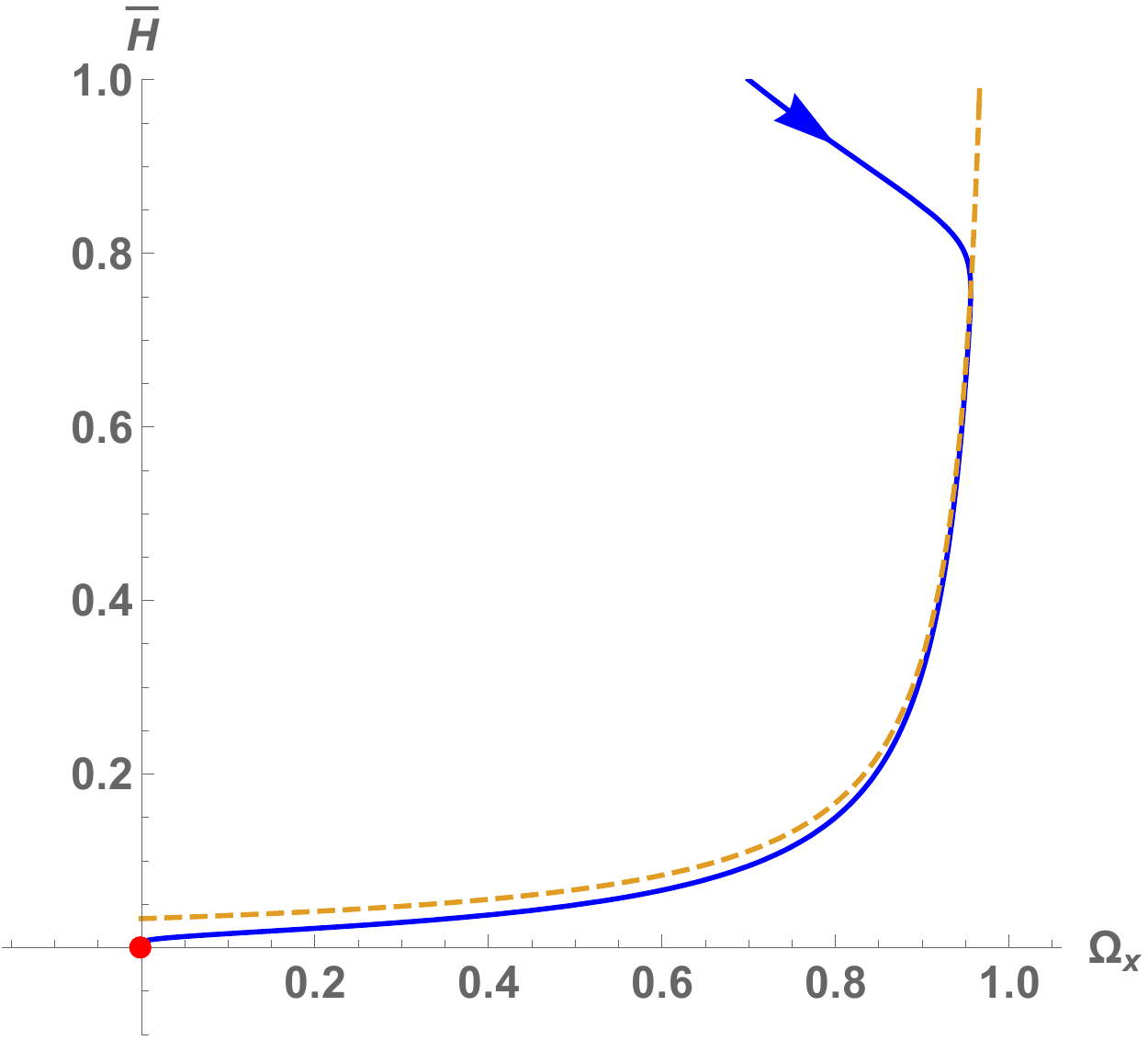}
\caption{Phase portrait for IVS1 for $\bar\Gamma = -0.1$. The blue line is the orbit with initial conditions $\Omega_x=0.7$ and
$\bar{H}=1$. The dashed line represents the points with $\Omega_x'=0$.}
\label{portrait-IVS1}
\end{figure*}

  To understand analytically this behavior 
  we use the system (\ref{dynamical1}), replacing $\Omega_c$ by $\Omega_x$, which for our model is given by 
  \begin{eqnarray}\label{2}\left\{\begin{array}{ccc}
    {\Omega}'_x   &=& (3\bar{H}(1-\Omega_x)+\bar{\Gamma}) \Omega_x, \\
    \bar{H}'  & =&-\frac{3}{2}\bar{H}^2(1-\Omega_x),
  \end{array}\right.
  \end{eqnarray}
  which has only the fixed point $(\Omega_x=0,\bar{H}=0)$. 
  Unfortunately, the linearization does not decide because the determinant of the linear matrix is zero at the fixed point.
  But, depicting the curve 
$3\bar{H}(1-\Omega_x)+\bar{\Gamma}=0$ (it is  $\Omega_x'=0$) 
and the field  (\ref{2}) in the plane $(\Omega_x, \bar{H})$, 
one can see in Fig. \ref{portrait-IVS1} that this point is an attractor for $\bar{\Gamma}<0$.

  \subsection{IVS2}
  
\begin{figure*}
    \centering
    \includegraphics[width=1.0\textwidth,trim={0 3.5cm 0 3.5cm},clip]{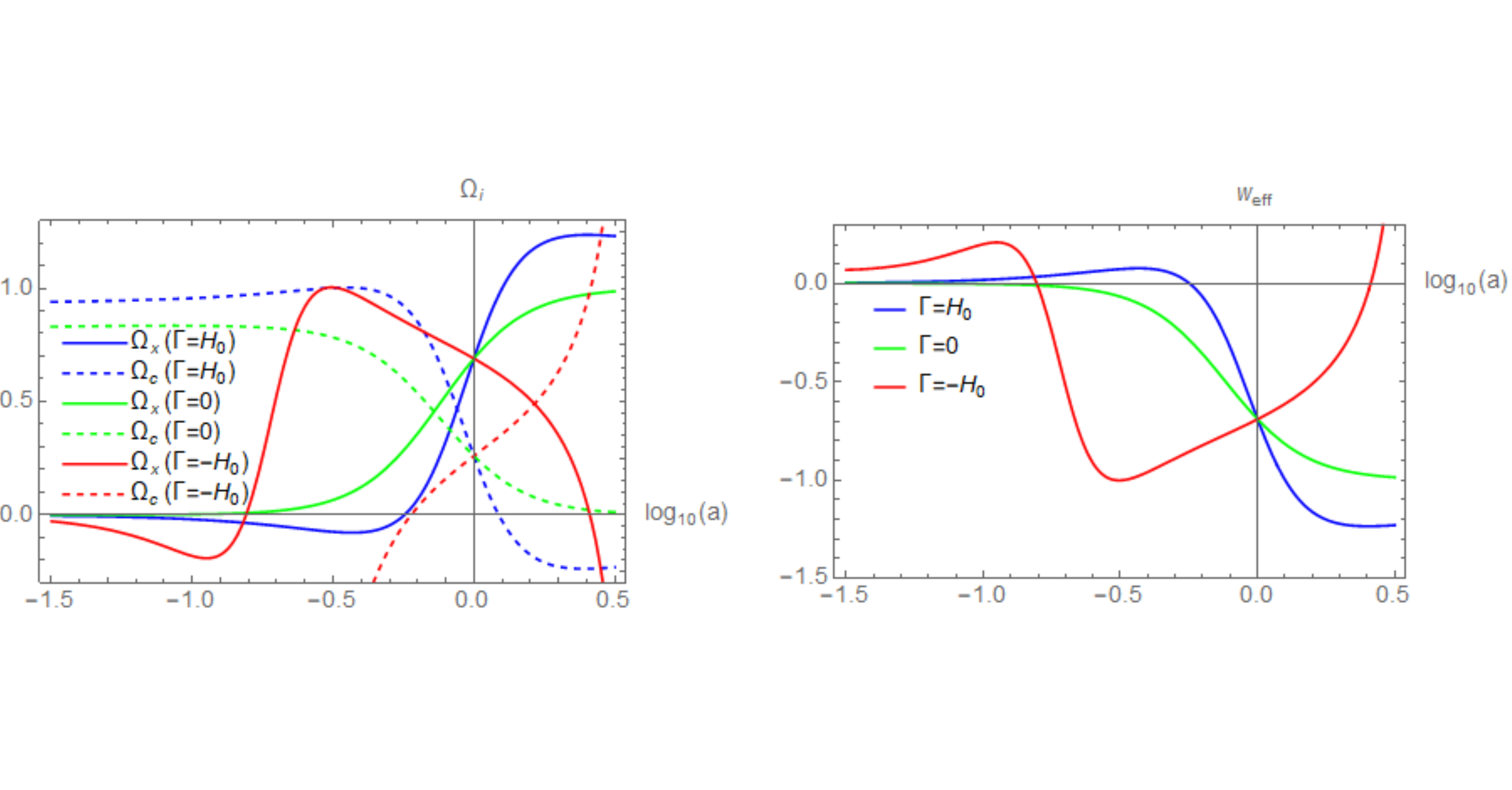}
    \caption{We show the evolution of the density parameters for dark matter and dark energy (left plot) and the total equation of state for different values of the dimensionless coupling parameter, namely  $\Gamma/H_0= 1 $, $\Gamma/H_0=0$ and $\Gamma/H_0=-1$ for the interaction function (\ref{model2}). While drawing the plots, we have fixed $H_0=68$ km/sec/Mpc, $\Omega_{c0}=0.26$, $\Omega_{b0} = 0.0499$, $\Omega_{r0} = 0.0001$ and  $\Omega_{x0}=0.69$.}
    \label{fig:IVS2}
\end{figure*}

\begin{figure*}
\includegraphics[width=0.45\textwidth]{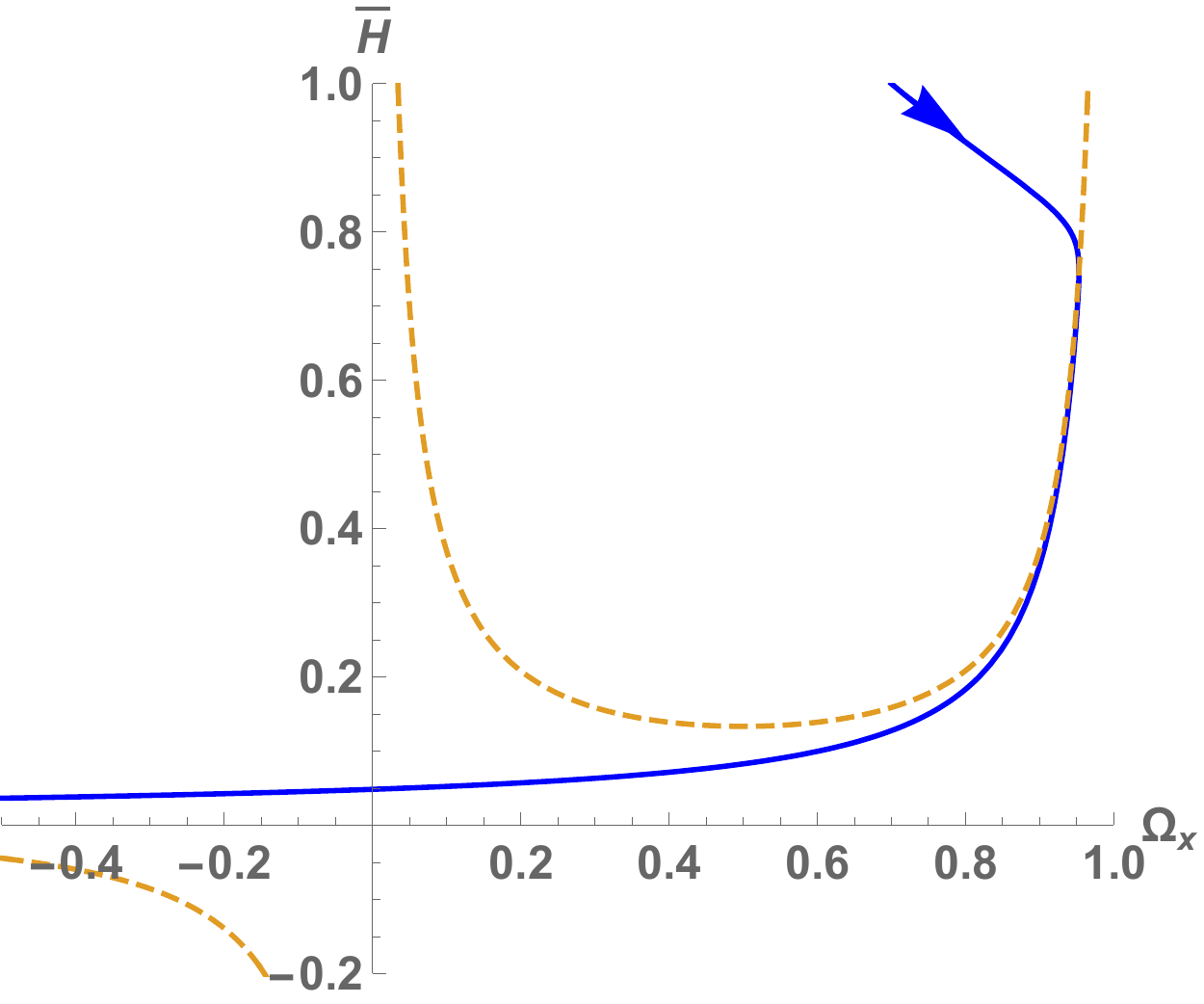}
\caption{Phase portrait for IVS2 for $\bar\Gamma=-0.1$. The blue line is the orbit with initial conditions $\Omega_x=0.7$ and 
$\bar{H}=1$. The dashed line represents the points with $\Omega_x'=0$.}
\label{portrait-IVS2}
\end{figure*}

  This model is completely non viable for $\Gamma/H_0>0$ because before the present time $\rho_x$ gets negative and after the present time $\rho_c$ has negative values (see Fig. \ref{fig:IVS2}). On the contrary, for $\Gamma/H_0<0$ at very late times, $\rho_c$ dominates once again, but looking at (\ref{dynamical1}) and 
  changing $\Omega_c$ by $\Omega_x$ one has
   \begin{eqnarray}\label{3}\left\{\begin{array}{ccc}
    {\Omega}'_x   &=& 3\bar{H}(1-\Omega_x)\Omega_x+\bar{\Gamma}, \\
    \bar{H}'  & =&-\frac{3}{2}\bar{H}^2(1-\Omega_x).
  \end{array}\right.
  \end{eqnarray}
  And we can see in Fig. \ref{portrait-IVS2}, from the plot of the vector field in the plane $(\Omega_x,\bar{H})$,  that the orbits cross the axis $\bar{H}=0$ from the first to the second quadrant, meaning that $\Omega_x$ becomes negative at late times, and consequently, the model is non viable for any value of $\Gamma$ different from zero.

  \begin{figure*}
    \centering
    \includegraphics[width=1.0\textwidth,trim={0 3.5cm 0 3.5cm},clip]{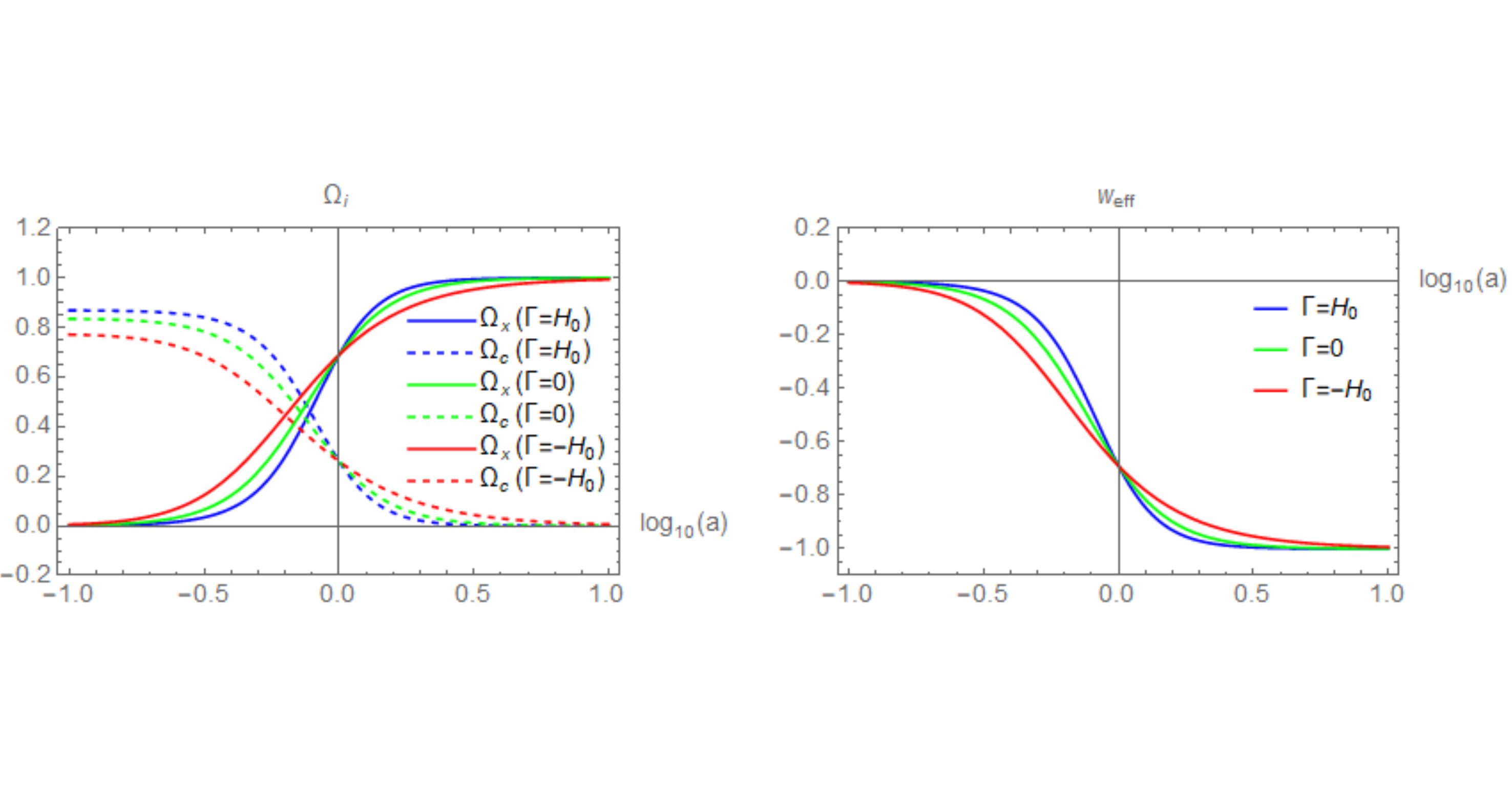}
    \caption{We show the evolution of the density parameters for dark matter and dark energy (left plot) and the total equation of state for different values of the dimensionless coupling parameter, namely  $\Gamma/H_0= 1$, $\Gamma/H_0=0$ and $\Gamma/H_0=-1$ for the interaction function (\ref{model3}). While drawing the plots, we have fixed $H_0=68$ km/sec/Mpc, $\Omega_{c0}=0.26$, $\Omega_{b0} = 0.0499$, $\Omega_{r0} = 0.0001$ and  $\Omega_{x0}=0.69$.  }
    \label{fig:IVS3}
\end{figure*}

 \subsection{IVS3} 
 
 For this model all cases are viable, and asymptotically at late times, for positive and negative values of $\Gamma/H_0$, the dynamics are the same as in the non-interacting case, that means, it converges to the fixed point $(\Omega_c=0, \Omega_x=1)$.

 To see that, we continue with the system (\ref{dynamical1})
 \begin{eqnarray}\left\{\begin{array}{ccc}
    {\Omega}'_c   &=& -(3\bar{H}+\bar{\Gamma})\Omega_c(1-\Omega_c) \\
    \bar{H}'  & =&-\frac{3}{2}\bar{H}^2\Omega_c,
  \end{array}\right.
  \end{eqnarray}  
  which has as fixed points
   $(1,0)$  and the line $\Omega_c=0$. From the above two equations, we obtain the following solvable equation,
  \begin{eqnarray}
  \frac{d\Omega_c}{d\bar{H}}=\frac{2(3\bar{H}+\bar{\Gamma})}{3\bar{H}^2}(1-\Omega_c),
  \end{eqnarray}
  which by integrating, we get that
   \begin{eqnarray}
   1-\Omega_c(\bar{H})=\frac{\Omega_{x,0}}{\bar{H}^2} \exp\left[\frac{2\bar\Gamma}{3\bar{H}}(1-\bar{H})\right].
   \end{eqnarray}
  
So, we have to find the zeros of $\Omega_c(\bar{H})$. When $\bar{\Gamma}$ is positive, $\Omega_c$ has only a zero for $\bar{H}$ between $0$ and $1$, which is obviously an attractor, meaning that at late times our universe enters into an accelerating phase with $w_{\rm eff}=-1$. On the contrary, when $\bar{\Gamma}$ is negative, $\Omega_c$ has two zeros and we have numerically found that the greater one, namely $\bar{H}_*$, is between $\bar{H}=0.515$ and $\bar{H}=1$. Since the system starts with $\bar{H}=1$, this means that $\bar{H}_*$ is an attractor, and thus, as in the positive case, the universe accelerates forever with $w_{\rm eff}=-1$ (see Fig. \ref{fig:IVS3}).

\section{Observational data, statistical method and results}
\label{sec-data+analyses}

In this section we describe the observational data used to confront the interaction scenarios and present their results. 

\begin{itemize}

\item \textbf{Cosmic Microwave Background (CMB):} The data from CMB are very poweful to constrain the dark energy models. In this article we have used the latest CMB data from the Planck 2018 final data release. To be specific, we make use of the CMB temperature and polarization angular power spectra {\rm plikTTTEEE+lowl+lowE} \cite{Aghanim:2018eyx,Aghanim:2019ame}.

\item \textbf{Baryon acoustic oscillations (BAO)}:  We also use the BAO data from different astronomical surveys  6dFGS~\cite{Beutler:2011hx}, SDSS-MGS~\cite{Ross:2014qpa}, and 
BOSS DR12~\cite{Alam:2016hwk}. The use of BAO is essential because the degeneracies appearing in the constraints of the associated cosmological parameters obtained from the CMB data alone.
    
\end{itemize}

We consider the seven dimensional parameter space for each interacting scenario where the free parameters are as follows: the baryon energy density $\Omega_bh^2$; the energy density of the cold dark matter $\Omega_{c}h^2$; the optical depth to the reionization $\tau$; the spectral index of the primordial scalar perturbations $n_s$; the ratio of the sound horizon at decoupling to the angular diameter distance to the last scattering $100 \theta_{MC}$; its amplitude $A_s$, and lastly the dimensionless coupling parameter $\Gamma/H_0$.  The flat priors on the first six parameters ($\Omega_bh^2$, $\Omega_{c}h^2$, $\tau$, $n_s$, $100 \theta_{MC}$, $A_s$) are shown in 
Table~\ref{tab:priors} while the prior on the remaining parameter $\Gamma/H_0$ has been fixed from the analyses presented in section \ref{sec-3}. 

\begin{table}
\begin{center}
\begin{tabular}{|c|c|c}
\hline
Parameter                    & Prior\\
\hline 
$\Omega_{b} h^2$             & $[0.005,0.1]$\\
$\Omega_{c} h^2$             & $[0.001,0.99]$\\
$\tau$                       & $[0.01,0.8]$\\
$n_s$                        & $[0.8,1.2]$\\
$100\theta_{MC}$             & $[0.5,10]$\\ 
$\log[10^{10}A_{s}]$         & $[1.6,3.9]$\\
$\Gamma/H_0$                 & Different for different IVS models\\
\hline
\end{tabular}
\end{center}
\caption{Flat priors assumed on various cosmological parameters.}
\label{tab:priors}
\end{table}

Now, to constrain all the interacting scenarios, we modified the 
Markov Chain Monte Carlo code \texttt{CosmoMC} (see the details here \url{http://cosmologist.info/cosmomc/})~\cite{Lewis:2002ah,Lewis:1999bs}, an excellent cosmological package equipped with Planck  2018 likelihood~\cite{Aghanim:2019ame}. The package has a convergence diagnostic following the Gelman-Rubin statistics quantified through $R-1$ ~\cite{Gelman-Rubin}. 

In the following we describe the main results extracted from all the interacting scenarios that are physically viable as described in section \ref{sec-3}.

\subsection{IVS0}

\begin{figure}
\includegraphics[width=0.48\textwidth]{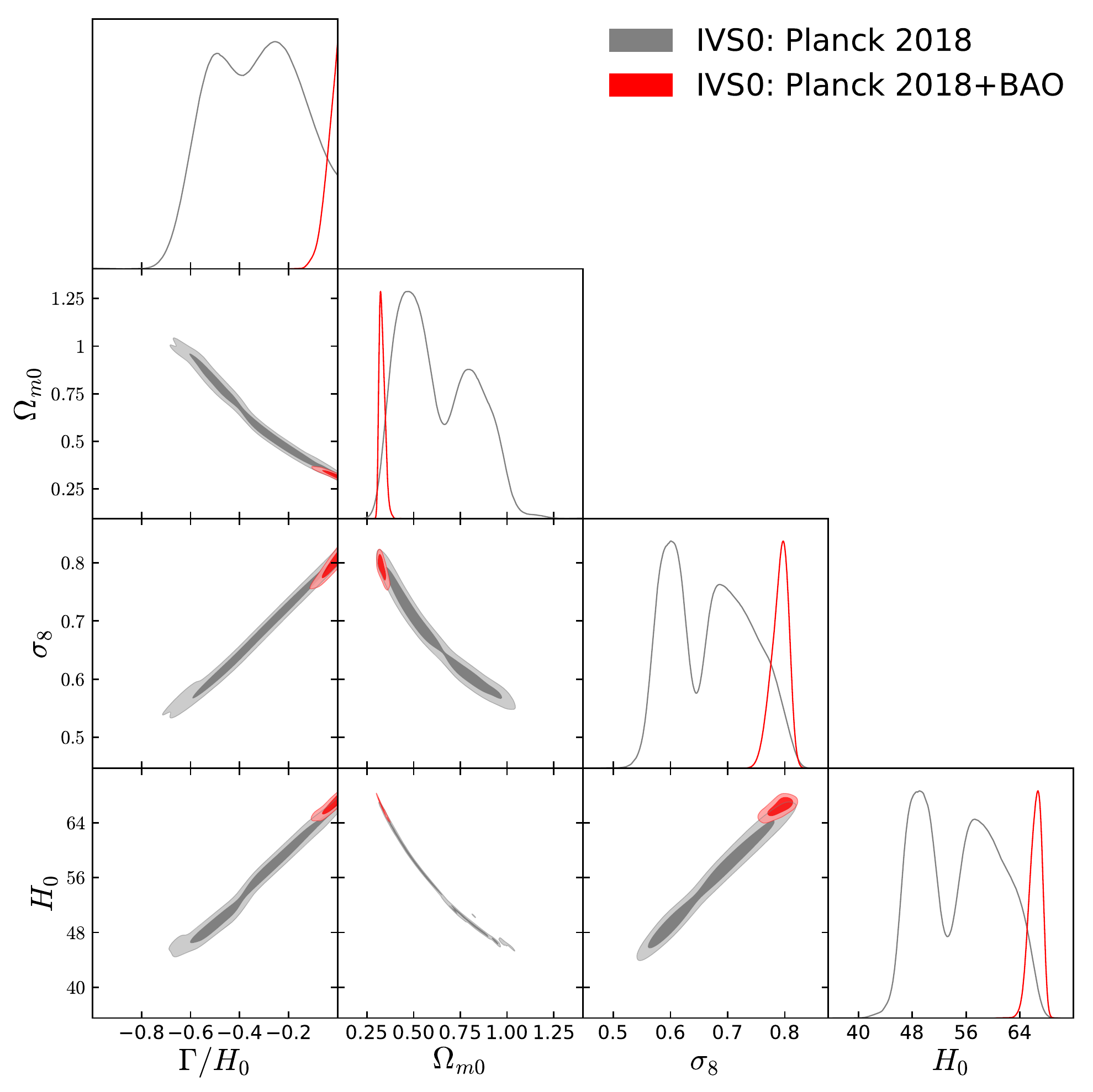}
\caption{We show the one dimensional marginalized posterior distributions of some key parameters  together with the two dimensional joint contours of the IVS0 scenario considering Planck 2018 and Planck 2018+BAO datasets. }
\label{fig:ivs0}
\end{figure}

As explained in section \ref{sec-3}, for this interaction model the dark energy density becomes negative in the past when $\Gamma > 0$, that means, the positive prior on $\Gamma$ leads to unphysical properties of the cosmological parameters. Therefore, during the observational analysis we have imposed negative prior on $\Gamma/H_0 \in (-1, 0)$. The results are summarized in Table \ref{tab:ivs0} and the corresponding graphical variations are shown in Fig. \ref{fig:ivs0}, where we have specifically shown the one dimensional posterior distributions of some key parameters of this scenario as well as the two dimensional contour plots.    

Concerning the analyses with Planck 2018 and Planck 2018+BAO, let us  note that the cosmological bounds obtained from Planck 2018  shown in Table~\ref{tab:ivs0} are not acceptable because in this case we see the  bimodal distribution (see Fig. \ref{fig:IVS0} for Planck 2018 data only). Hence, the analysis with Planck 2018 is not powerful enough to distinguish one of the two peaks, and hence, the constraints are not reliable. The appearance of the bimodal distribution is not  new in the cosmological context because in some other cosmological models, see for instance Ref. \cite{Benaoum:2020qsi}, this has been found.  Therefore, from the constraint on the dimensionless coupling parameter $\Gamma/H_0$, one cannot say whether the interacting picture is really preferred or not, even if $\Gamma/H_0$  is nonzero at more than 68\% CL ($\Gamma/H_0 =   -0.34_{- 0.21}^{+ 0.18}$ at 95\% CL for Planck 2018 data only). This result is not surprising as also explored recently in \cite{DiValentino:2020leo}. 

The inclusion of BAO with Planck 2018 makes the constraints more reliable and this is in agreement with a non-interacting scenario within 68\% CL reporting $\Gamma/H_0 = -0.034_{-  0.052}^{+  0.034}$ (Planck 2018+BAO).   However, the Hubble constant as we can see from Table \ref{tab:ivs0} is low ($H_0 = 66.33_{-    0.78}^{+    1.09}$ at 68\% CL) compared to the Planck's $\Lambda$CDM based estimation \cite{Aghanim:2018eyx}. This is due to the strong anti-correlation existing between $\Gamma/H_0$ and $H_0$ itself.   

Further, in Table \ref{tab:ivs0} we have also compared the fitting of the model with respect to the non-interacting $\Lambda$CDM model by presenting the values of $\Delta \chi^2$. We exclude the $\Delta \chi^2$ computed for Planck 2018 data alone since, as argued above, due to the bimodal distribution appearing in this case (see Fig. \ref{fig:IVS0}), the constraints are not reliable and hence $\Delta \chi^2$ is also not reliable. For the Planck 2018+BAO dataset, we see that $\Delta \chi^2  =-0.97$ indicates a mild preference of the $\Lambda$CDM model over this interaction model.

\begingroup                                                 
\squeezetable                                                                                                                   
\begin{center}                                                                                                                  
\begin{table}                                                                                                                   
\begin{tabular}{cccccccc}                                                                                                            
\hline\hline                                                                                                                    
Parameters & Planck 2018 & Planck 2018+BAO \\ \hline

$\Omega_c h^2$ &  $    0.157_{-    0.019-    0.036}^{+    0.022+    0.034}$ & $    0.1226_{-    0.0034-    0.0049}^{+    0.0018+    0.0058}$\\

$\Omega_b h^2$ & $    0.02236_{-    0.00015-    0.00031}^{+    0.00016+    0.00031}$ & $    0.02242_{-    0.00015-    0.00028}^{+    0.00014+    0.00029}$ \\

$100\theta_{MC}$ & $    1.03876_{-    0.00108-    0.00162}^{+    0.00094+    0.00176}$ & $    1.04062_{-    0.00031-    0.00065}^{+    0.00032+    0.00065}$  \\

$\tau$ & $    0.052_{-    0.0075-    0.015}^{+    0.0079+    0.015}$ &  $    0.057_{-    0.0089-    0.017}^{+    0.0073+    0.017}$ \\

$n_s$ & $    0.9743_{-    0.0046-    0.0091}^{+    0.0046+    0.0095}$  &  $    0.9758_{-    0.0040-    0.0077}^{+    0.0041+    0.0075}$ \\

${\rm{ln}}(10^{10} A_s)$ & $    3.049_{-    0.017-    0.033}^{+    0.017+    0.032}$ & $    3.057_{-    0.018-    0.031}^{+    0.015+    0.036}$   \\

$\Gamma/H_0$ & $   -0.34_{-    0.21-    0.29}^{+    0.18+    0.34}$ & $   -0.034_{-    0.0082-    0.052}^{+    0.034+    0.034}$  \\

$\Omega_{m0}$ & $    0.63_{-    0.27-    0.34}^{+    0.24+    0.38}$  & $    0.331_{-    0.019-    0.027}^{+    0.011+    0.031}$  \\


$H_0$ & $   55.15_{-    8.51-   10.60}^{+    7.20+   11.11}$  & $   66.33_{-    0.78-    1.82}^{+    1.09+    1.68}$  \\


\hline 
$\chi^2$ & $2770.316$ & $2780.66$ \\
$\Delta \chi^2$ & $2.85$  & $-0.97$ \\
\hline\hline                                                                                                        
\end{tabular}                 
\caption{68\% and 95\% CL constraints on various free and derived parameters of the interacting scenario IVS0 corresponding to the interaction function $Q= \Gamma \rho_c $ (IVS0) using Planck 2018 and Planck 2018+BAO datasets. We also show the $\chi^2$  values for the best-fit parameters and $\Delta \chi^2 = \chi^2$ ($\Lambda$CDM) $-$ $\chi^2$ (IVS0). A negative value of $\Delta \chi^2$  infers that $\Lambda$CDM is preferred over the interacting scenario while the positive value of $\Delta \chi^2 $ infers the opposite.  }
\label{tab:ivs0}                                                                                                   
\end{table}                                                                                                                     
\end{center}                                                                                                                    
\endgroup                                                                                                                    

\subsection{IVS1}
\label{sec-results-IVS1}
\begin{figure}
\includegraphics[width=0.48\textwidth]{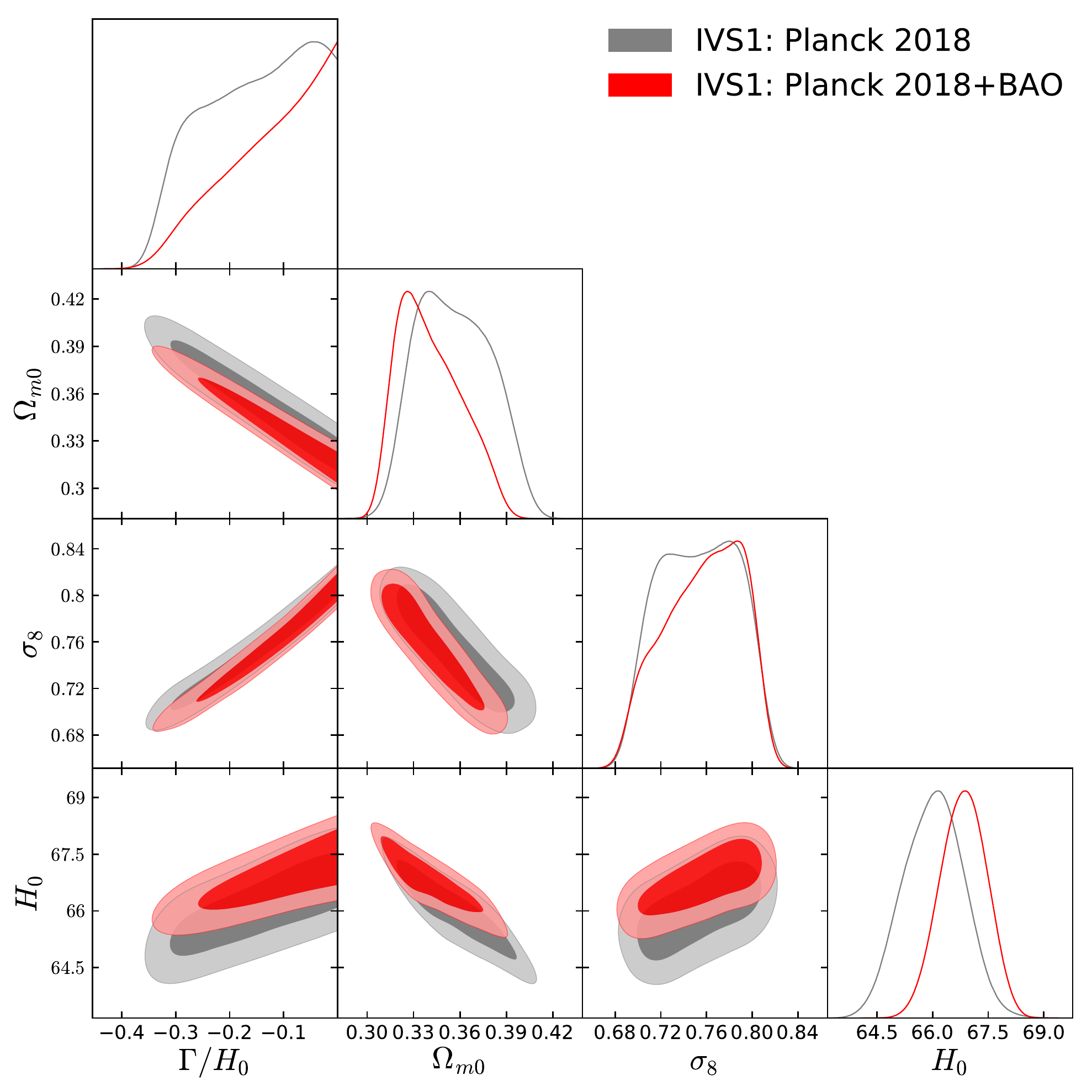}
\includegraphics[width=0.48\textwidth]{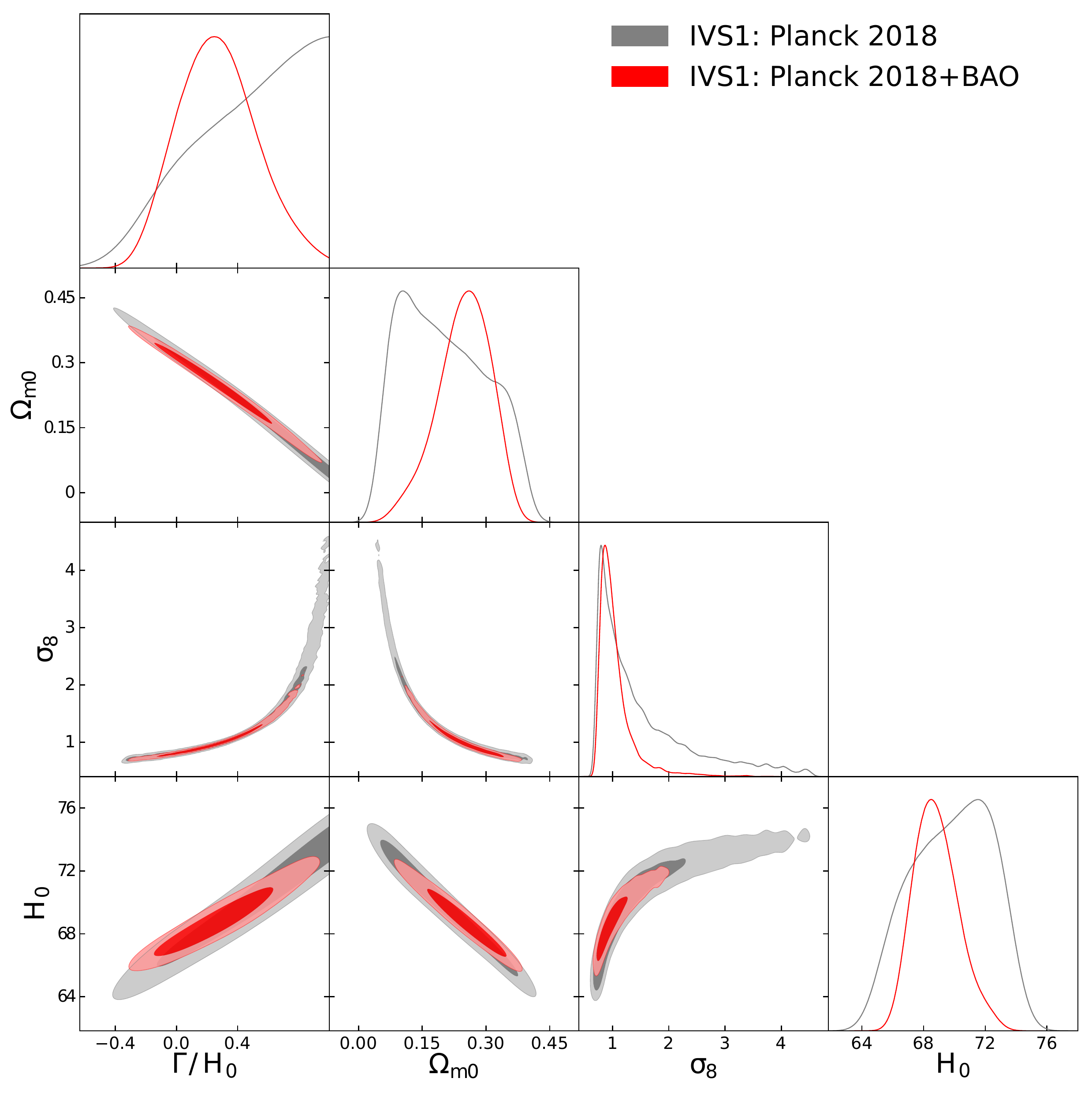}
\caption{We show the one dimensional marginalized posterior distributions of some key parameters  together with the two dimensional joint contours of the IVS1 scenario considering Planck 2018 and Planck 2018+BAO datasets. The upper plot corresponds to the case where $\Gamma/H_0$ is varying in the region $(-1, 0)$ while the lower plot corresponds to the case where $\Gamma/H_0$ is varying in the region $(-1, 1)$.    }
\label{fig:ivs1}
\end{figure}

As shown in section \ref{sec-3}, this model exhibits negative dark energy density in the future for positive values of the coupling parameter $\Gamma$ while from early time to 
the present time, the model does not lead to any unphysical behavior for both signs of the coupling parameter $\Gamma$. 
Therefore, in the observational analyses we have considered two separate cases: (i) when $\Gamma/H_0$ is in the negative region, that means $\Gamma/H_0 \in (-1, 0)$ and secondly (ii) when $\Gamma/H_0$ is  
varying in $(-1, 1)$. The consideration of both positive and negative values of $\Gamma/H_0$ is justified for the observational fittings since
we are dealing with the cosmological datasets available up to the present time and the model is viable for both signs of $\Gamma/H_0$ up to the present time.  With such considerations, we perform the observational fittings of this scenario and the results are summarized in Table \ref{tab:ivs1} and Fig. \ref{fig:ivs1}.  Let us describe the findings in detail for both the priors on $\Gamma/H_0$. 

The upper half of Table \ref{tab:ivs1} and the upper plot of Fig. \ref{fig:ivs1} correspond to the results for $\Gamma/H_0 \in (-1, 0)$. One can clearly notice that $\Gamma/H_0$ is consistent to the non-interacting scenario within 68\% CL for both Planck 2018 and Planck 2018+BAO datasets. The estimations of the Hubble constant for both the datasets are slightly lower than the Planck's estimations (within $\Lambda$CDM model) \cite{Aghanim:2018eyx} and due to the existing anti-correlation between $H_0$ and $\Omega_{m0}$, we can see larger values of $\Omega_{m0}$ obtained from both the datasets. Finally, from the $\Delta \chi^2$ values displayed at the end of the upper half of Table \ref{tab:ivs1}, one can see that even if for Planck 2018 alone, the interaction model seems to be preferred ($\Delta \chi^2 = 0.82$), however, the presence of BAO data makes this conclusion reversed ($\Delta \chi^2 = -0.58$ for Planck 2018+BAO).

We now discuss the results for the remaining case, that means when $\Gamma/H_0$ is varying in $(-1, 1)$. 
As we can see from the lower half of Table \ref{tab:ivs1}, the dimensionless coupling parameter $\Gamma/H_0$ assumes nonzero value for both Planck 2018,
which remains nonzero within 68\% CL but within 95\% CL the zero value of $\Gamma/H_0$ is allowed. However, such non-zero estimation of $\Gamma/H_0$ does not actually infer the presence of an interaction in the dark sector since it could be fake, as explored recently in  \cite{DiValentino:2020leo}. 
The Hubble constant assumes higher values compared to the $\Lambda$CDM based Planck 2018 estimation \cite{Aghanim:2018eyx} with significantly higher error bars. 
When BAO data are added to Planck 2018, the magnitude of the coupling parameter decreases and, within 68\% CL, the non-interacting picture becomes consistent. But the Hubble constant is slightly higher than the $\Lambda$CDM based Planck 2018 estimation \cite{Aghanim:2018eyx} although not significantly. Anyway, the notable point is that here the error bars on $H_0$ are higher compared to the $\Lambda$CDM based Planck 2018 estimation \cite{Aghanim:2018eyx}. And that is why such higher error bars weakly alleviate the $H_0$ tension and this is purely due to the higher error bars.  

Finally, we computed the $\Delta \chi^2$ values for both Planck 2018 and Planck 2018+BAO which are shown at the end of the lower half of  Table \ref{tab:ivs1}. We find that for Planck 2018 alone, $\Lambda$CDM is indeed preferred over this interaction model ($\Delta \chi^2  = -0.24$), while for Planck 2018+BAO we have a different indication where the IVS1 seems to be preferred ($\Delta \chi^2 = 1.27$) over the $\Lambda$CDM model. Let us note that we have an exactly different conclusion in this case compared to the previous case with $\Gamma/H_0 \in (-1, 0)$ (see the values of $\Delta \chi^2$ summarized at the end of the first half of the Table \ref{tab:ivs1}). So, we can see that the sign of $\Gamma/H_0$ is indeed important because this sign controls the direction of the energy flow between the dark sectors.

\begingroup                                                                                                                     
\squeezetable                                                                                                                   
\begin{center}                                                                                                                  
\begin{table}                                                                                                                   
\begin{tabular}{cccccccccccc}                                                                                                            
\hline                                               

& {\it When $\Gamma/H_0$ is considered} &  {\it to be varying in $(-1, 0)$}\\
\hline 
Parameters & Planck 2018 & Planck 2018+BAO \\ \hline

$\Omega_c h^2$ & $    0.1321_{-    0.0095-    0.0122}^{+    0.0067+    0.0130}$ & $    0.1290_{-    0.0094-    0.0110}^{+    0.0047+    0.0135}$ \\

$\Omega_b h^2$ & $    0.02230_{-    0.00015-    0.00028}^{+    0.00015+    0.00031}$ & $    0.02241_{-    0.00014-    0.00027}^{+    0.00014+    0.00027}$ \\

$100\theta_{MC}$ & $    1.03996_{-    0.00052-    0.00092}^{+    0.00052+    0.00092}$ & $    1.04025_{-    0.00045-    0.00092}^{+    0.00053+    0.00085}$  \\

$\tau$ & $    0.054_{-    0.0082-    0.016}^{+    0.0078+    0.016}$ & $    0.056_{-    0.0077-    0.016}^{+    0.0079+    0.016}$  \\

$n_s$ & $    0.9722_{-    0.0042-    0.0082}^{+    0.0043+    0.0080}$ & $    0.9756_{-    0.0037-    0.0076}^{+    0.0037+    0.0073}$ \\

${\rm{ln}}(10^{10} A_s)$ & $    3.055_{-    0.016-    0.034}^{+    0.016+    0.032}$ & $    3.056_{-    0.016-    0.033}^{+    0.016+    0.034}$ \\

$\Gamma/H_0$ & $   -0.15_{-    0.054-    0.16}^{+    0.15+    0.15}$ & $   -0.13_{-    0.040-    0.17}^{+    0.13+    0.13}$ \\

$\Omega_{m0}$ & $    0.356_{-    0.030-    0.042}^{+    0.023+    0.045}$ & $    0.341_{-    0.027-    0.035}^{+    0.015+    0.041}$  \\


$H_0$ & $   66.02_{-    0.84-    1.55}^{+    0.81+    1.50}$ & $   66.83_{-    0.64-    1.23}^{+    0.64+    1.17}$ \\


\hline 
$\chi^2$ & $2772.35$ & $2780.274$ \\
$\Delta \chi^2$ & $0.82$ & $-0.58$ \\
\hline
& {\it When $\Gamma/H_0$ is considered} &  {\it to be varying in $(-1, 1)$} \\
\hline 
Parameters & Planck 2018 & Planck 2018+BAO \\ \hline

$\Omega_c h^2$ & $    0.0719_{-    0.0605-    0.0709}^{+    0.0344+    0.0656}$ & $    0.0934_{-    0.0190-    0.0544}^{+    0.0318+    0.0479}$ \\

$\Omega_b h^2$ & $    0.02230_{-    0.00015-    0.00030}^{+    0.00015+    0.00030}$ & $    0.02234_{-    0.00015-    0.00029}^{+    0.00015+    0.00029}$   \\

$100\theta_{MC}$ & $    1.04384_{-    0.00395-    0.00506}^{+    0.00259+    0.00558}$ & $    1.04232_{-    0.00198-    0.00300}^{+    0.00106+    0.00356}$ \\

$\tau$ & $    0.054_{-    0.0080-    0.015}^{+    0.0074+    0.015}$ & $    0.055_{-    0.0078-    0.015}^{+    0.0076+    0.016}$  \\

$n_s$ & $    0.9721_{-    0.0041-    0.0084}^{+    0.0042+    0.0080}$ & $    0.9735_{-    0.0041-    0.0078}^{+    0.0040+    0.0079}$   \\

${\rm{ln}}(10^{10} A_s)$ & $    3.055_{-    0.015-    0.031}^{+    0.015+    0.031}$ & $    3.056_{-    0.017-    0.031}^{+    0.016+    0.033}$  \\

$\Gamma/H_0$ & $    0.449_{-    0.177-    0.667}^{+    0.551+    0.551}$ & $    0.269_{-    0.298-    0.505}^{+    0.240+    0.526}$ \\

$\Omega_{m0}$ & $    0.201_{-    0.142-    0.171}^{+    0.082+    0.193}$ & $    0.248_{-    0.054-    0.133}^{+    0.076+    0.122}$  \\


$H_0$ & $   69.86_{-    2.40-    5.08}^{+    3.45+    4.65}$ & $   68.83_{-    1.73-    2.81}^{+    1.26+    3.09}$  \\

\hline 
$\chi^2$ & $2773.408$ & $2778.42$ \\
$\Delta \chi^2$ & $-0.24$ & $1.27$ \\
\hline\hline                                                                 
\end{tabular}                                                                                                                 
             
\caption{68\% and 95\% CL constraints on various free and derived parameters of the interacting scenario IVS1 corresponding to the interaction function $Q= \Gamma \rho_x $ (IVS1) using Planck 2018 and Planck 2018+BAO datasets.  We also show the $\chi^2$  values for the best-fit parameters and $\Delta \chi^2 = \chi^2$ ($\Lambda$CDM) $-$ $\chi^2$ (IVS1). A negative value of $\Delta \chi^2$  infers that $\Lambda$CDM is preferred over the interacting scenario while the positive value of $\Delta \chi^2 $ infers the opposite. The upper half of the table corresponds to the constraints when $\Gamma/H_0$ is restricted in $(-1, 0)$ while the lower half of the table corresponds to the constraints when $\Gamma/H_0$ is freely varying in $(-1, 1)$.}
\label{tab:ivs1}                                                                                                  
\end{table}                                                                                                                     
\end{center}                                                                                                                    
\endgroup

\subsection{IVS3}

\begin{figure}
\includegraphics[width=0.48\textwidth]{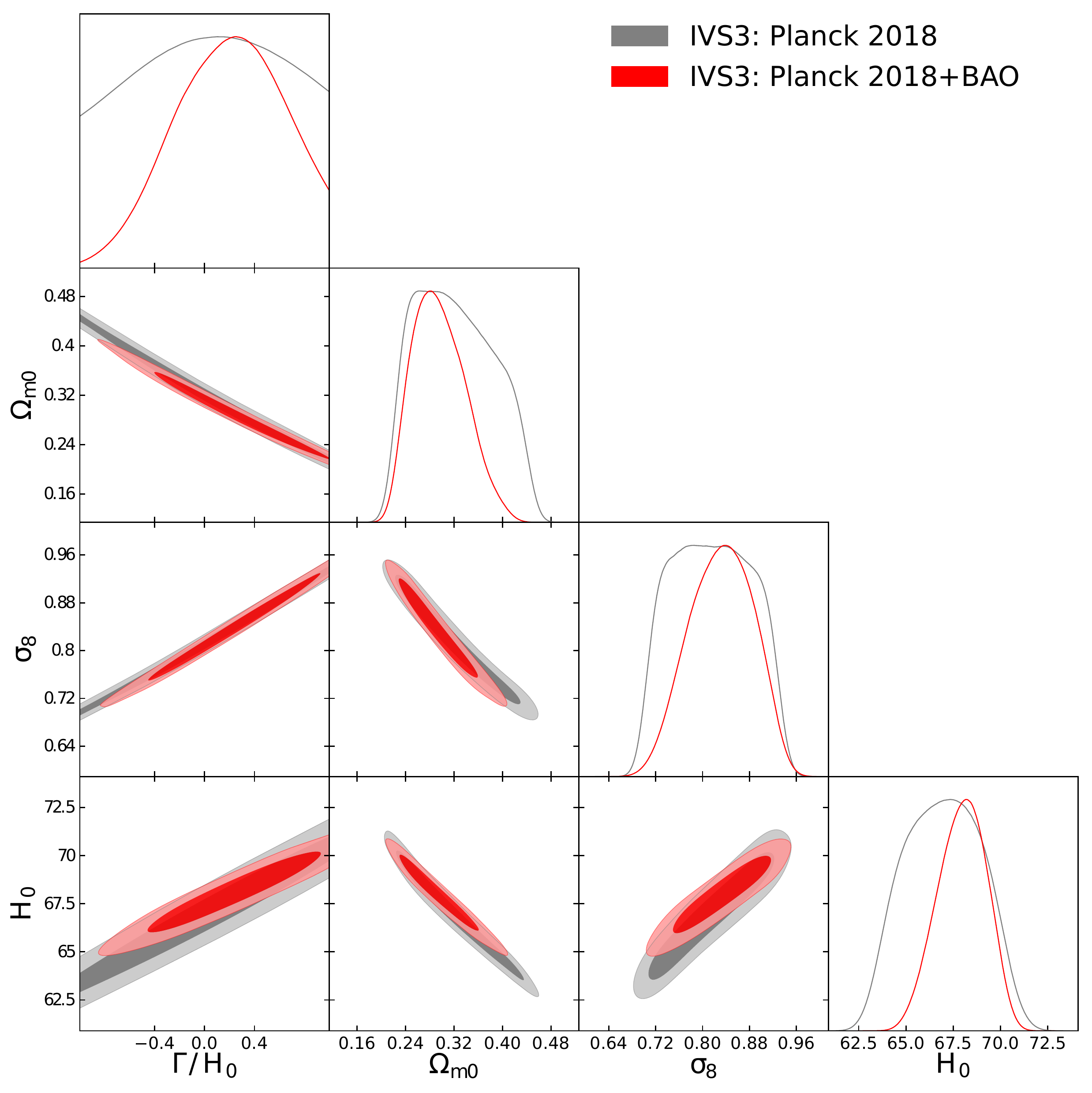}
\caption{We show the one dimensional marginalized posterior distributions of some key parameters  together with the two dimensional joint contours of the IVS3 scenario considering Planck 2018 and Planck 2018+BAO datasets.   }
\label{fig:ivs3}
\end{figure}

This is the only model in this series of interaction models which does not show any irregularities in the cosmological parameters unlike other interaction models (see section \ref{sec-3}). That means this model works fine for any value of the coupling parameter. 
We have constrained this interaction scenario using the same cosmological datasets as for the other remaining scenarios. The results are summarized in Table \ref{tab:ivs3} and in Fig. \ref{fig:ivs3}.

We can see that for both Planck 2018 and Planck 2018+BAO datasets, the non-interacting scenario is easily recovered within the 68\% CL as one can see from the estimations of the dimensionless coupling parameters:
$ \Gamma/H_0 = 0.024_{-    1.024}^{+    0.976}$ (68\% CL, Planck 2018) and 
$ \Gamma/H_0 =   0.188_{- 0.427}^{+    0.478}$ (68\% CL, Planck 2018+BAO). Therefore, the evidence of an interaction is not indicated within this interacting scenario. Additionally, concerning the estimations of the Hubble constant, we see that it assumes lower values compared to the minimal $\Lambda$CDM based Planck's measurements \cite{Aghanim:2018eyx}. 

In a similar fashion, we have also computed the values of $\Delta \chi^2$ for this model with respect to the $\Lambda$CDM scenario and we see $\Delta \chi^2  = 0.36$ (for Planck 2018) and   $\Delta \chi^2  = -1.38$ (for Planck 2018+BAO). As we can see, even though Planck 2018 data prefer mildly the interaction in the dark sector, Planck 2018+BAO dataset says differently, that means for this datatset $\Lambda$CDM is favored.

\begingroup                                                                                                                     
\squeezetable                                                                                                                   
\begin{center}                                                                                                                  
\begin{table}                                                                                                                   
\begin{tabular}{ccccccccccc}                                                                                                            
\hline\hline                                                                                                                    
Parameters & Planck 2018 & Planck 2018+BAO \\ \hline

$\Omega_c h^2$ & $    0.1196_{-    0.0252-    0.0358}^{+    0.0209+    0.0367}$ & $    0.1128_{-    0.0176-    0.0283}^{+    0.0147+    0.0291}$  \\

$\Omega_b h^2$ & $    0.02230_{-    0.00015-    0.00029}^{+    0.00015+    0.00030}$ & 
 $    0.02237_{-    0.00014-    0.00030}^{+    0.00016+    0.00028}$  \\

$100\theta_{MC}$ & $    1.04068_{-    0.00141-    0.00197}^{+    0.00120+    0.00205}$ & $    1.04114_{-    0.00086-    0.00167}^{+    0.00097+    0.00163}$   \\

$\tau$ & $    0.055_{-    0.0084-    0.015}^{+    0.0077+    0.016}$ & $    0.056_{-    0.0082-    0.015}^{+    0.0075+    0.016}$   \\

$n_s$ & $    0.9722_{-    0.0042-    0.0083}^{+    0.0042+    0.0084}$ & $    0.9741_{-    0.0041-    0.0079}^{+    0.0040+    0.0078}$   \\

${\rm{ln}}(10^{10} A_s)$ & $    3.056_{-    0.017-    0.032}^{+    0.016+    0.033}$ & 
$    3.056_{-    0.017-    0.031}^{+    0.015+    0.034}$  \\

$\Gamma/H_0$ & $    0.024_{-    1.024-    1.024}^{+    0.976+    0.976}$ & $    0.188_{-    0.427-    0.716}^{+    0.478+    0.812}$  \\

$\Omega_{m0}$ & $    0.321_{-    0.087-    0.112}^{+    0.056+    0.122}$ & $    0.295_{-    0.054-    0.081}^{+    0.039+    0.087}$  \\


$H_0$ & $   66.98_{-    2.23-    3.59}^{+    2.20+    3.55}$ & $   67.98_{-    1.28-    2.58}^{+    1.46+    2.49}$  \\


\hline 
$\chi^2$ & $2772.812$ & $2781.068$ \\
$\Delta \chi^2$ & $0.36$  & $-1.38$ \\
\hline\hline                                                       
\end{tabular}                                                                                                                 
\caption{68\% and 95\% CL constraints on various free and derived parameters of the interacting scenario IVS3 corresponding to the interaction function $Q= \Gamma \rho_c \rho_x /(\rho_c+\rho_x) $ using Planck 2018 and Planck 2018+BAO datasets. We also show the $\chi^2$  values for the best-fit parameters and $\Delta \chi^2 = \chi^2$ ($\Lambda$CDM) $-$ $\chi^2$ (IVS3). A negative value of $\Delta \chi^2$  infers that $\Lambda$CDM is preferred over the interacting scenario while the positive value of $\Delta \chi^2 $ infers the opposite. }
\label{tab:ivs3}                                                                                                   
\end{table}                                                 
\end{center}                                                                                                            
\endgroup 

\section{Summary and concluding remarks}
\label{sec-discuss}

In this article we studied some non-gravitational interaction models between dark matter and dark energy where the interaction function does not allow the  explicit 
presence of the Hubble factor. Some authors argue that the interaction between dark matter and dark energy should be a local phenomenon, although it is not yet clearly understood whether the presence of the dark energy in the early times can be ruled out or not and consequently the interaction in the early universe or in the intermediate matter dominated phase can be allowed as well. Additionally, on the other hand, if the expansion of the universe suddenly stops, then the interaction rates allowing Hubble rate explicitly should immediately vanish, meaning that the interaction is dependent only on the expansion of the universe. As the interaction is mainly governed by the properties of dark matter and dark energy, therefore, such models are sometimes criticized.    
While the nature of the dark sectors is completely unknown, whether the interaction rate should contain the Hubble expansion factor or not is very hard to comment. Hence, the motivation of the current work has been to investigate the interaction models of the form $Q = \mathcal{F} (\rho_c, \rho_x)$
 containing no $H$ explicitly just to impose the theoretical priors on them and finally to examine their observational fitness.  
 
 We have investigated four interaction models, namely Model 0: $Q =  \Gamma \rho_c$,  Model 1: $Q = \Gamma \rho_x$, Model 2: $Q =  \Gamma (\rho_c +\rho_x)$, and Model 3: $Q = \Gamma \rho_c \rho_x (\rho_c + \rho_x)^{-1}$,  
 which are obtained from a very general interaction model $Q =  \Gamma\rho _{c}^{\alpha }\rho _{x}^{1-\alpha -\beta }(\rho _{c}+\rho_{x})^{\beta },$
where $\alpha$ and $\beta$ are real numbers and $\Gamma $ is the coupling parameter with ${\rm dimension}(\Gamma) = {\rm dimension} (H_0)$. 

We have studied all four models in order to check their viability and imposed the theoretical bounds on them in terms of the dimensionless coupling parameter $\Gamma/H_0$, and we have shown that the coupling parameter plays a very crucial role in this case. We have performed the dynamical system analysis for each model (see section \ref{sec-3}). 
We have found that all four models are not able to offer viable interacting scenarios since the energy density of either dark matter or dark energy could be negative either in the past or in the future depending on the sign of the coupling parameter. More elaborately, we have seen that, for $\Gamma >0$, the dark energy density becomes negative in the past within IVS0 ($Q =  \Gamma \rho_c$) scenario, but for $\Gamma<0$ this model works fine and the energy densities remain positive. For IVS1 ($Q =  \Gamma \rho_x$), we have seen that the energy densities of the dark sectors remain positive for all $\Gamma$ in the past, but in the future the dark matter density becomes negative for $\Gamma > 0$.   
Moreover, within the context of IVS2, the energy density of the dark energy becomes negative in the past for both $\Gamma > 0$ and $\Gamma <0$.  That means IVS2 does not lead to any physically acceptable scenario neither in the past nor in the future.  But IVS3 is viable for all $\Gamma$, that means, the energy densities of both dark matter and dark energy remain positive throughout the evolution.

Considering the theoretical restrictions on the interaction models, we have constrained the viable interacting scenarios
using the CMB data from Planck 2018 data release and Planck 2018+BAO dataset. The results are shown in Tables \ref{tab:ivs0}, \ref{tab:ivs1} and \ref{tab:ivs3} and the corresponding plots are shown in Figs. \ref{fig:ivs0}, \ref{fig:ivs1} and \ref{fig:ivs3}.  The inclusion of BAO data to CMB alone is motivated to break the degeneracies amongst the parameters. Thus, the presence of BAO is therefore very important to avoid the fake claim about the indication of an interacting scenario that can be obtained from CMB alone dataset \cite{DiValentino:2020leo}. In fact, the indication of an interaction has been found in IVS0 ($\Gamma/H_0 \neq 0$ at more than 68\% CL), while, as one may note, the constraints are not reliable due to the bimodal distribution  (see Fig. \ref{fig:ivs0}) and therefore we are mainly concerned with the results from Planck 2018+BAO dataset for all the scenarios.  We note that for IVS1, we have considered two separate cases in the observational analysis, namely for $\Gamma/H_0 \in (-1, 0)$ and for $\Gamma/H_0 \in (-1, 1)$ as explained in section \ref{sec-results-IVS1}.  
From the analyses, we find that for all three models, namely IVS0, IVS1, IVS3, Planck 2018+BAO dataset indicates a non-interacting model of the universe together with a preference for the $\Lambda$CDM model over the IVS models (with the exception of IVS1 for $\Gamma/H_0 \in (-1, 1)$) quantified through  $\Delta \chi^2$ analysis.

\section{Acknowledgments}
We thank the referee for her/his useful comments that helped us to improve the article. WY is supported by the National Natural Science Foundation of China under Grants No. 11705079 and No. 11647153, and Liaoning Revitalization Talents Program under Grant no. XLYC1907098.  SP has been supported by the Mathematical Research Impact-Centric Support Scheme  (MATRICS), File No. MTR/2018/000940, given by the Science and Engineering Research Board (SERB), Govt. of India. JdH  has been supported by MINECO (Spain) grant MTM2017-84214-C2-1-P, and  in part by the Catalan Government 2017-SGR-247.

\section{In memory of Prof. John D. Barrow}

We dedicate this work to Prof. John D. Barrow, who passed away at the end of the last year making everyone of us very sad.  This paper and also a follow up paper that will be posted after some time are greatly connected with Prof. Barrow. After we (WY and SP) wrote an article on interacting dark energy \cite{Yang:2017zjs} with Prof. Barrow, he suggested us to examine the interaction models without the Hubble expansion factor $H$.  We were mainly concerned about the analyses of such models at the level of perturbations and consequently the modifications of our codes. Several months elapsed during the modifications of the codes, running of the chains, and finally we came up with this version, but Prof. Barrow left us. We shall always miss Prof. Barrow for his immense contributions in cosmology and for his greatness. 


\end{document}